\documentclass[conference,compsoc]{IEEEtran}

\ifCLASSOPTIONcompsoc
  \usepackage[nocompress]{cite}
\else
  \usepackage{cite}
\fi

\ifCLASSINFOpdf
\else
\fi
\begin{document}

\title{Apply SOA Paradigms in Cyber-Physical System to Enhance Interoperability: \\ State-of-the-Art Review}

\author{\IEEEauthorblockN{Su Zhang}
\IEEEauthorblockA{School of Electronics Engineering and Computer Science\\
Peking University, Beijing, China\\
Email: samsuzhang@pku.edu.cn}}

\maketitle

\begin{abstract}
\\The Cyber-Physical System (CPS) is considered to be the next generation of intelligent industrial automation systems that integrate computing, communication and control technologies. In CPS, the interoperability requirements between devices and processes are growing rapidly. However, the diversity and heterogeneity of technologies and standards between information technology (IT) and operational technology (OT) in industrial systems pose significant challenges.
Service-Oriented Architecture (SOA) provides a software design pattern in which software components are connected only through messaging, which is consistent with the fundamentals of modularity and communication in distributed automation systems. SOA is considered to be a key technology for improving interoperability in next-generation industrial automation systems.
This paper introduces the latest research advances in applying SOA paradigms to CPS to improve interoperability, including: Open Platform Communications Unified Architecture (OPC UA) and related extensions, implementing SOA at the manufacturing device level, integrating legacy systems in SOA-based automation systems, Plug-and-Play in SOA-based automation systems, dynamic configuration in SOA-based automation systems and the combination of SOA with Multi-Agent Systems and Evolvable Production Systems.


\end{abstract}

\IEEEpeerreviewmaketitle

\section{Introduction}
Today, information technology and communication technology are rapidly evolving, and the convergence of the physical world and the online world has created a new form of production system, the Cyber-Physical System (CPS) \cite{lee2015cyber}.

CPS emphasizes the real-time, dynamic information feedback and circulation process between the physical world and the information world. It deeply integrates various information technologies: sensors, embedded computing, cloud computing, network communication, and software engineering. These highly collaborative and autonomous information technologies enable production applications to be monitored independently, intelligently, dynamically and systematically, which changes the traits of the physical world.The goal of CPS is to achieve deep integration between the information system and the physical world, as well as between various information systems.

In the industrial field, the production system accepts and processes the information (perception) feedback from the physical world while manipulating the mechanical equipment. The “uploading” and “downloading” of the information in the production control have a high degree of real-time synergy. CPS is designed to meet this characteristic. Therefore, CPS is considered to be the next-generation intelligent industrial automation system integrating computing, communication and control technology \cite{rajkumar2010cyber}.

It has attracted great attention from academia, industry and the government, and has been widely used in manufacturing because of its potential benefits to society, economy, energy and the environment, especially manufacturing.

However, the diversity and heterogeneity of technologies and standards between information technology (IT) and operational technology (OT) in industrial systems pose significant challenges.Service-Oriented Architecture (SOA) provides a software design pattern in which software components are connected only through messaging, which is consistent with the fundamentals of modularity and communication in distributed automation systems. SOA is considered to be a key technology for improving interoperability in next-generation industrial automation systems.

This paper introduces the latest research advances in applying SOA paradigms to CPS to improve interoperability, including: Open Platform Communications Unified Architecture (OPC UA) and related extensions, implementing SOA at the manufacturing device level, integrating legacy systems in SOA-based automation systems, Plug-and-Play in SOA-based automation systems, dynamic configuration in SOA-based automation systems and the combination of SOA with Multi-Agent Systems and Evolvable Production Systems.

The remainder of this paper is organized as follows. In Section II, the interoperability issues in CPS are discussed. In Section III, many of the latest research advances in applying SOA paradigms to CPS to improve interoperability are introduced. Section V summarizes the whole paper and makes a vision.

\section{Interoperability Issues in CPS}
Significant advances in communication standards and protocols have led to more connected devices and systems \cite{nagorny2012service}, where a large number of heterogeneous devices are often integrated into more complex systems to enhance the functionality of industrial manufacturing systems.

The biggest challenge facing a new generation of industrial systems may not be security, but interoperability and data portability \cite{gruner2016restful}, as the connected devices involved in the new business model grow exponentially.

In the various concepts of Industry 4.0 \cite{bechtold2014industry}, Industrial Internet \cite{lin2015industrial} and Cyber-Physical Production System (CPPS) \cite{monostori2015cyber}, control execution and management of decisions that monitor and coordinate physical processes are made through creating a digital representation of the physical world.

The main requirements that must be met by a new generation of manufacturing companies and industrial systems include the following \cite{jammes2005service}: dynamic integration capabilities across enterprises, cross-enterprise collaboration, support for heterogeneous but interoperable hardware and software environments, achieving flexibility through adaptability and reconfigurability, achieving scalability by adding resources without interrupting operations, fault tolerance and the capacity of returning from failure.

In this process, the following objectives must be achieved to ensure the integration of next-generation industrial systems: cross-component structural connections, semantic interoperability of components and systems, and architectures that are independent of open vendors \cite{nagorny2012service}.

As a result, interoperability requirements between devices and processes are growing rapidly across all systems from the factory to the enterprise. However, the diversity and heterogeneity of technology and standards in industrial systems (especially at the factory floor) pose significant challenges \cite{jirkovsky2017understanding}.

In addition, the nature of the technology and communication standards used in the factory floor is different from the technology used in enterprise systems \cite{panetto2008enterprise}. This problem of interoperability between information technology (IT) and operational technology (OT) is also known in the industry as IT-OT convergence \cite{harp2014ot}.

\section{Apply SOA Paradigms to CPS to Improve Interoperability}
\subsection{SOA in The Field of General Computing}
The concept of Service Oriented Architecture (SOA) has been introduced in the field of general computing to facilitate the creation of distributed network computer systems.

Several well-known SOA principles have been defined in the field of software engineering \cite{erl1900service}. The primary goal of SOA is to increase the flexibility, interoperability, and level of abstraction of software components. The two key principles of SOA are loose coupling and discoverability. The SOA definition logic must be encapsulated into the service, and the service can only be accessed through messages. Loosely coupled software components ensure future expansion to a variety of platforms and future technologies. The complex logic hidden in the service provides an abstraction of the system-level overview. Services and messages are formally defined by contracts. The service contract is registered in the service repository so that other services can discover and invoke the service.

The loose coupling introduced by SOA ensures interoperability between platforms, regardless of their hardware and software architecture. Finally, by exchanging service contracts, each service registered in the service repository can be discovered and invoked from other services, increasing the reusability of the program.

\subsection{Apply SOA in Industrial Automation Systems}
The goal of CPPS is to shift from mass production to mass customization in manufacturing systems based on service-oriented network physical components \cite{wahlster2016industrie}.

SOA provides a software design pattern in which software components are only connected via messaging. This design pattern is consistent with the basic principles of distributed automation systems: modularity and communication \cite{vyatkin2013software} \cite{zhu2010optimizing}. SOA is considered to be a key technology for improving interoperability in next-generation industrial automation systems \cite{jammes2005service} \cite{van2010flexibility}.

F. Jammes et al. outlined the opportunities and challenges of the next generation of embedded devices, applications and services, and envisioned the future direction of intelligent device networks based on service-oriented advanced protocols \cite{jammes2005service}.

V. Van Tan provides advice on applying SOA paradigms to automated systems implemented through Web services technologies for adoption at all levels of the automation hierarchy \cite{van2010flexibility}.

T. Cucinotta et al. describe an enhanced SOA that can meet the real-time and quality of service (QoS) of industrial automation real-time applications \cite{cucinotta2009real}.

\subsection{OPC UA and Related Extensions}
Open Platform Communication Unified Architecture (OPC UA) is a platform-independent SOA-based industrial system standard whose specifications are organized into several documents related with concepts, security models, address space models, services, information models, mappings, configuration files, Data access, alerts and conditions, procedures and history access. OPC UA offers the possibility to extend interoperability \cite{mahnke2009opc} \cite{melik2012towards}.

OPC UA supports both information transfer and modeling, and its information modeling framework supports the integration of information models into protocols, enabling them to transfer information in vendor-specific formats. J. Virta et al. proposed a method for combining OPC UA with SOA-based middleware and the ISA-88/95 standard to integrate Manufacturing Execution Systems (MES) and Process Control Systems (PCS) in a batch management environment\cite{virta2010soa}. ISA-95 has become one of the supporting information models specified by OPC UA \cite{brandl2013opc}.

In OPC UA, the startup cost of building an initial information model is very high. AutomationML (Automated Markup Language) is an upcoming standard series (IEC 62714) for describing production plants or plant components. R. Henssen et al. proposed a method to simplify the creation of an OPC UA information model based on existing AutomationML data by examining the analogy between the AutomationML and the OPC UA information model \cite{henssen2014interoperability}.

Representational State Transfer (REST) is a broad architectural style for distributed applications. REST recommends using a fixed set of service interfaces to transport heterogeneous resource representations instead of defining custom interfaces for individual applications. S. Gruner et al. explored the communication advantages and system design advantages of service-oriented software architecture including RESTful services in industrial environments, and proposed a RESTful extension to OPC UA binary protocol, which can significantly reduce communication overhead during system interaction \cite{gruner2016restful}.

Devices Profile for Web Services (DPWS) specifies how Web Service technology can be applied to embedded devices, enabling devices to dynamically join the network and discover other devices on the network and the services they provide. M. J. A. G. Izaguirre et al. proposed an implementation solution that uses complex event handlers to manage heterogeneous events coming in from OPC UA and DPWS to improve the interoperability between OPC UA and DPWS in distributed systems for monitoring and monitoring plant floors \cite{izaguirre2011opc}.

\subsection{Implement SOA at the Manufacturing Device Level}
Vertical integration using a Web services-based service-oriented architecture in the manufacturing equipment layer of industrial enterprises can significantly improve the operability and flexibility of inter-manufacturing layer devices, enabling seamless integration of business software and manufacturing tasks.

M. Mathes et al. introduced a SOAP engine (SOAP4PLC) for programmable logic controllers that enables business applications to invoke manufacturing tasks running on programmable logic controllers (PLCs) as Web services to drive the employment of Web Services in the manufacturing layer \cite{mathes2009soap4plc}.

C. Stoidner et al. proposed an extension based on the SOAP4PLC engine to enable it to invoke Web services-based business applications from PLC applications \cite{stoidner2010invoking}.

\subsection{Integrate Legacy Systems in SOA-based Automation Systems}
Existing work on SOA-based automation systems typically uses (Web) services in device-level directly, while there are currently incompatible existing devices in mass production systems.

O.Givehchi et al. proposed an interoperability layer between network (IT) and physical (OT) systems that can provide minimally modified migration method to achieve connectivity and interoperability for legacy systems by applying the Common Information Model (CIM) based on the ISA95 industry standard \cite{givehchi2017interoperability}.

S. Feldhorst et al. proposed a thin abstraction layer that provides the technical functions of an industrial plant as a reusable service that can be arranged in a control hierarchy or at a higher level of workflow to integrate legacy systems into device-level SOA \cite{feldhorst2009integration}.

A. Girbea et al. proposed a service-oriented architecture that uses a set of OPC UA servers to model information from the device level to address integration issues in legacy systems \cite{girbea2014design}.

F. Tao et al. proposed an Industrial IoT Hub (IIHub), which includes a Custom Access Module (CA-Module), Access Hub (A-Hub) and Local Service Pool (LSP), with the capacity of discovering and connecting with Internet/Intranet, service encapsulation, and intelligent analytics and precision management, a flexible set of CA modules can be configured or programmed to connect heterogeneous physical manufacturing resources \cite{tao2018iihub}.

\subsection{Plug-and-Play in SOA-based Automation Systems}
In modern factories, new trends in customer demand are forcing manufacturers to move from mass production to small-volume custom products. There are a wide variety of products for each product type, so the production line must be adjusted frequently and needs to be more resistant to anomalies that may occur during manufacturing execution, and Plug-and-Play devices are becoming more and more important.

W. Dai et al. proposed a Plug-and-Play service component approach based on Web Services Description Language (WSDL) and IEC 61499 functional blocks \cite{dai2016enabling}.

G. Cândido et al. implemented an SOA-based infrastructure that facilitates the process of device exchange in an industrial automation environment by matching semantics and device semantics during device replacement \cite{candido2013enhancing}.

V. Jirkovsky et al. proposed a Plug-and-Play solution that uses OPC Unified Architecture and Semantic Web technologies to achieve integration at all levels, allowing self-describing devices to be Plug-and-Play. Insert into a larger system \cite{jirkovsky2018towards}.

The knowledge-driven approach introduces new possibilities for the efficient operation of the system, but it also requires a lot of effort to describe the knowledge about the system. S. Iarovyi et al. proposed a method based on combining a knowledge-driven approach with a service-oriented architecture to automatically leverage data from manufacturing equipment to facilitate knowledge base collections to reduce the cost of introducing the Open Knowledge-Driven Manufacturing Execution System (OKD-MES) into the industry combining a knowledge-driven approach with a service-oriented architecture, and evaluated the possibility of extracting the required semantic data from the description provided by the intelligent service-oriented manufacturing equipment \cite{iarovyi2015representation}.

\subsection{Dynamic Configuration in an SOA-based Automation System}
Future networked embedded systems will be complex deployments that are deeply integrated into the environment they are monitoring. They must react to user and environmental events, which may require modifying their structure to handle changing conditions. In many areas, including industrial environments, this modification of the system architecture will need to be done in real time.

J. Cecilio et al. proposed a method in the form of software architecture for enabling embedded devices to be remotely (re)configured in a sensor and actuator network with flexible computing capabilities. An intermediate computing layer was designed to hide the embedded different hardware implementations of device network applications, which provides the ability to remotely (re)configure distributed systems for embedded devices \cite{cecilio2014architecture}.

W. Dai et al. proposed a method for applying SOA to a distributed automation system based on the IEC 61499 standard at the device level. The method implements a formal definition of SOA based on the IEC 61499 standard, designed to increase the flexibility of distributed automation systems and reduce development and integration costs. It can dynamically create or delete any element in a function block without interrupting the normal execution in SOA-based IEC 61499 runtime \cite{dai2015bridging}.

\subsection{The Combination of SOA and Other Concepts}
The Multi-Agent System is a system that intelligently and flexibly responds to changes in operating conditions and the needs of surrounding processes. K. Nagorny et al. propose a collaborative industrial manufacturing automation architecture that combines SOA with a Multi-Agent System designed to support structural connectivity and functional interoperability between devices and systems in a customizable service-oriented enterprise architecture, which consider the standard enterprise architecture from a functional perspective and help manage and control networked intelligent automation components in a distributed manufacturing environment \cite{nagorny2012service}.

Devices have a lifecycle, from initial setup and deployment to system lifecycle monitoring and evolution, each device needs to be considered and easily accessible. By linking Evolvable Production System (EPS) and SOA paradigms, G. Cândido et al. proposed a common architecture solution that supports interoperability at different stages of the device lifecycle (such as device setup, control, management, monitoring, and diagnostics) \cite{candido2011service} .

\section{Conclusion}
In the Cyber-Physical System (CPS), the interoperability requirements between devices and processes are growing rapidly. Service-Oriented Architecture (SOA) provides a loosely coupled design pattern to increase the flexibility, interoperability and level of abstraction of system components. Based on SOA, factory floor equipment can be integrated with modern enterprise services in a service-oriented manner to help integrate Information Technology (IT) and Operational Technology (OT) in CPS.

We have introduced the latest research progress in the application of SOA in CPS, including: OPC UA and related extensions, implementing SOA in the manufacturing equipment layer, integrating legacy systems in SOA-based automation systems, Plug-and-Play in SOA-based automation systems, dynamic configuration in SOA-based automation systems and the combination of SOA with Multi-Agent Systems and Evolvable Production Systems.

In addition, the combination of services and the implementation of reliable services in SOA-based industrial automation systems has yet to be carried out.
In the field of general computing, there has been a lot of research work on these directions, which is worth learning.For example, in terms of service management, X. Liu et al. conducted a study on the automation service combination \cite{liu2008user} \cite{liu2010community}, Y. Ma et al. conducted a study on runtime service combination management \cite{ma2013model}, and G. Huang et al. conducted a study on model-based automated navigation and composition of complex service mashups \cite{huang2015model}; in terms of service reliability, G. Huang et al. studied the balance between the reliability and performance of services \cite{huang2006performance} and the optimization of dependable SOA based on reflective middleware \cite{huang2007soar}.

Furthermore, there is usually a clear trade-off between performance and flexibility: the higher the performance, the lower the flexibility that can be achieved. Introducing SOA in CPS brings additional performance overhead while increasing system flexibility and interoperability. Of course, different application scenarios have different real-time requirements for the automation system. As long as the overhead introduced by the SOA does not exceed the real-time constraints of the specific system, it will not have a negative impact on the system operation. At present, there is relatively little research work on reducing the performance overhead caused by the application of SOA. How to balance the flexibility and performance in service-oriented automation systems to meet the real-time constraints of specific systems needs further study.

\ifCLASSOPTIONcompsoc
\else
\fi

\bibliographystyle{IEEEtran}
\bibliography{reference}

\end{document}